\documentclass[aps,prl,preprint,superscriptaddress,floatfix,showpacs]{revtex4-1}
\usepackage{graphicx}
\usepackage{amsmath,amssymb,bm}
\begin{document}

\title{Unraveling the Salvinia paradox: design principles for submerged
superhydrophobicity}

\author{M. Amabili}
\affiliation{Sapienza Universit\`a di Roma, Dipartimento di Ingegneria
Meccanica e Aerospaziale, 00184 Rome, Italy}
\author{A. Giacomello}
\affiliation{Sapienza Universit\`a di Roma, Dipartimento di Ingegneria
Meccanica e Aerospaziale, 00184 Rome, Italy}
\author{S. Meloni}
\affiliation{Institute of Chemical Sciences and Engineering, \'Ecole
Polytechnique F\'ed\'erale de Lausanne, Lausanne, Switzerland}
\author{C.M. Casciola}
\affiliation{Sapienza Universit\`a di Roma, Dipartimento di Ingegneria
Meccanica e Aerospaziale, 00184 Rome, Italy}

\maketitle

%\textit{keywords}: 
%superhydrophobicity, Salvinia paradox, submerged surfaces, Cassie-Wenzel transition, rare event methods
%superhydrophobicity, Salvinia, submerged surfaces,
%Cassie-Wenzel transition, design criteria

%%%%%%%%%%%%%%%%%%%%%%%%%%%%%%%%%%%%%%%%%%%%%%%%

\begin{figure}
	\centering
	\includegraphics[width=7cm]{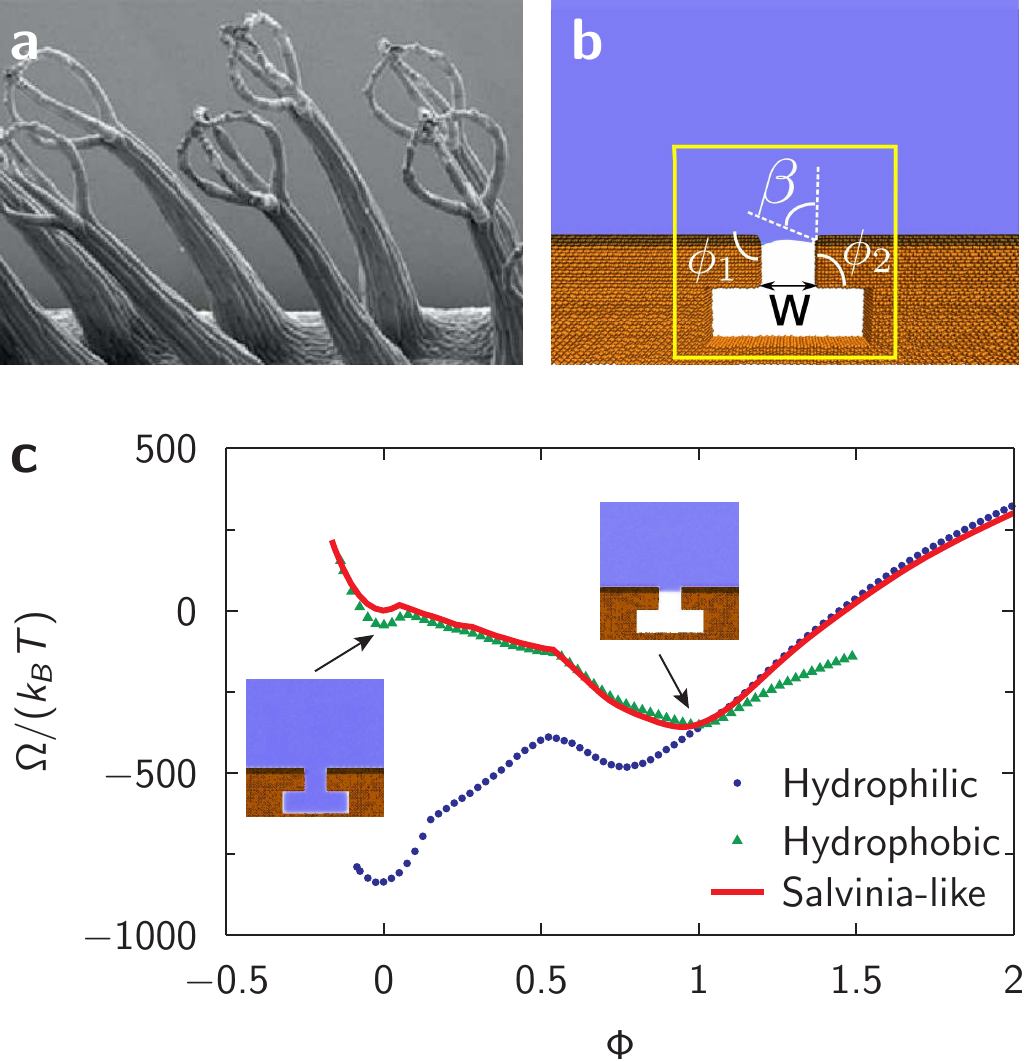}
	\caption{ 
		a) SEM micrograph of a Salvinia molesta leaf showing the  egg-beater
		hairs (adapted from	\cite{barthlott2010}).
		b) Atomistic Salvinia-like system used in RMD simulations. The fluid
		is represented in blue, the hydrophilic layer in dark brown,
		and the hydrophobic interior in light brown. The yellow lines define the box for the atom
		count collective variable $z$.\cite{giacomello2012langmuir}
		c) Free energy profiles at $\Delta P\approx 0$ as a function of the
		filling level $\Phi\equiv(z_W-z)/(z_W-z_C)$ for the hydrophobic
		system (blue dots), the hydrophilic one (green triangles), and the
		Salvinia-like one (red line).  
		$z_W$ ($z_C$) is computed in the Wenzel (Cassie hydrophobic) state, shown in the insets
		for the	Salvinia-like case. Thus, for the three chemistries, $\Phi = 0$ corresponds to the Wenzel
		state, $\Phi \sim 1$ to the Cassie state, and $\Phi > 1$ to a
		vapor bubble. 
	\label{fig:system}}
\end{figure}

\noindent Inspired by nature,\cite{barthlott1997,solga2007,koch2009} the study of
superhydrophobicity has flourished in the last two decades allowing for an
improved control of the wetting properties of surfaces of technological
interest.\cite{zhang2008,celia2013} In particular, submerged superhydrophobicity
is emerging as a means to reduce drag and prevent biofouling: such
applications require robust gas-trapping inside surface asperities.  
Here, we focus on the \emph{Salvinia paradox}\cite{barthlott2010}
assessing both via free energy atomistic simulations and via macroscopic
capillarity theory the role of the complex morphology
of this water fern (Figure~\ref{fig:system}a) in stabilizing an air
layer underwater. Our analysis shows that the
air-stabilizing mechanism of the Salvinia is in essence
determined by the pinning of the contact line and the
characteristic size of surface roughness. Simple design criteria for
stable submerged superhydrophobicity are devised, consolidating the
different approaches \cite{patankar2003,tuteja2007,butt2013}
within a common probabilistic framework.

Most applications of superhydrophobicity to date have concentrated on
drops deposited on surfaces, both statically and dynamically.
On the other hand, there is a growing interest in the properties of submerged
surfaces entrapping gas:\cite{checco2014,lv2014a,xu2014} in this
case, superhydrophobicity is a means to reduce the liquid-solid contact which,
in turn, diminishes drag \cite{joly2009,rothstein2010,gentili2014} and prevents
(bio)fouling.  Given their relevance for global industry and transportation,
even small improvements in the fuel efficiency and maintenance
costs of watercrafts and marine structures could have a significant
impact on society.\cite{schultz2011,ferrari2015} For submerged
applications the central question is the resistance and durability of
the gas pockets to pressure variations.  
In fact, the common-ground of
superhydrophobicity is the underlying ``suspended'' Cassie state in
which gas pockets are trapped within surface asperities.  Depending on
the external conditions, however, superhydrophobicity may break down in the fully
wet Wenzel state (Figure~\ref{fig:system}c).

The Lotus leaves have played a major role in inspiring the design of
drop-repellent surfaces.\cite{barthlott1997}  Moving towards submerged
applications requires a new inspiration: a promising candidate is 
the \emph{Salvinia molesta} (Figure~\ref{fig:system}a), because of its
superior vapor trapping capabilities.\cite{barthlott2010,cerman2009,hunt2011,mayser2014}

%%%%%%%%%%%%%%%%%%%%%%%%%%%%%%%%%%%%%%%%%%%%%%%%

The gas entrapped within surface asperities  can be either air or the
vapor phase coexisting with the liquid: albeit the partial pressure of
the other gases facilitates the Cassie state, their presence is not a
requirement for (meta)stable superhydrophobicity \cite{giacomello2013} 
(see \emph{Supporting Information} for additional details on the role of
dissolved gases). 
The entrapped gas may be lost through different mechanisms, analyzed in detail below, determining the failure of superhydrophobicity:
\begin{enumerate}
	\item mechanical destabilization of the Cassie state, e.g., due to an increase in the pressure (the \emph{spinodal} for the Cassie-Wenzel transition);
	\item \emph{thermally activated} Cassie-Wenzel transition; this process is much	slower than the spinodal one;
	\item gas loss in the liquid through \emph{thermally activated}
		nucleation of bubbles (at pressures below two-phase coexistence);
	\item \emph{spinodal} nucleation of bubbles (at pressures much below
		two-phase coexistence);
	\item air dissolution in the liquid.\footnote{This transient case, which is beyond the goal of the paper, is observed, e.g.,
in surface nanobubbles or when a surface is initially immersed in a liquid.}
\end{enumerate}

In order to rationalize these different scenarios and quantify the
``robustness'' of superhydrophobicity we compute the
probability to find the system in a generic macroscopic state $z$. This probability
$p(z; P,T)$ crucially depends on the thermodynamic conditions (pressure $P$
and temperature $T$) and is usually expressed in a logarithmic scale and
in units of the thermal energy -- the so-called Landau free energy
$\Omega$:
\begin{equation}
	p(z;P,T) = \exp{\left(-\frac{\Omega(z;P,T)}{k_B T} \right)} \;  ,
	\label{eq:landau}
\end{equation}
where $z$ is a variable characterizing the macroscopic state of
the system (here, the advancement of the Cassie-Wenzel transition) and $k_B$ is the
Boltzmann constant. 
When the free energy landscape $\Omega(z;P,T)$ is rugged, its local
minima correspond to highly probable  (meta)stable states, e.g.,
the Cassie and Wenzel states in the 1D landscape of Figure~\ref{fig:system}c.

The main pieces of information that can be extracted from $\Omega(z;P,T)$
are the free energy difference between any two states and the free energy barriers $\Delta \Omega^\dag$.  
In particular, the difference between two minima in the free energy measures the relative
probability of two (meta)stable states; for the case of the
Cassie-Wenzel transition, $\Omega_C -\Omega_W=-k_B T \ln(p_C/p_W)$. 
The maximum (\emph{t}ransition \emph{s}tate) separating two minima defines two free energy barriers, a ``forward'' and a
``backward'' one: $\Delta \Omega^\dag_{CW} \equiv
\Omega_\mathrm{ts}-\Omega_{C}$ and $\Delta \Omega^\dag_{WC} \equiv
\Omega_\mathrm{ts}-\Omega_{W}$, respectively.  According to the
transition state theory,\cite{eyring1935} the mean first passage time
between transitions from one minimum to the other depends exponentially on the
free energy barrier:
\begin{equation}
	\tau (P,T) = \tau_0 \exp{\left(\frac{\Delta\Omega^\dag(P,T)}{k_B T}
	\right)} \; .	
	\label{eq:rate}
\end{equation}
Summing up, the stability of a given state and the kinetics of the
transition to another state are ruled by the  free energy
landscape through Equation~\ref{eq:landau} and \ref{eq:rate}.

\emph{Rare event} techniques \cite{bonella2012} have been developed in
order to compute $p(z;P,T)$ on complex free energy landscapes overcoming
the extremely different timescales involved.
Here we employ \emph{restrained
molecular dynamics} (RMD, adapted from Ref.~\cite{TAMD}), which has
been shown to be effective in dealing with superhydrophobicity (see
Ref.~\cite{giacomello2012langmuir,giacomello2015} and
\emph{Supporting Information}). The advantage of using 
an atomistic description of the surface and of the liquid is that it relies on
minimal assumptions. Furthermore the dimensions of the simulated system
$\sim 5$~nm are sufficiently large so that it can be described in terms
of macroscopic capillarity \cite{Giacomello2012,checco2014} (see below);
comparing atomistic and macroscopic models makes it possible to draw
conclusions that are valid from the nano to the macro scale.

%%%%%%%%%%%%%%%%%%%%%%%%%%%%%%%%%%%%%%%%%%%%%%%%

The salient features of the Salvinia -- re-entrant geometry and
heterogeneous chemistry -- are captured in the simulations by a 
T-shaped cavity, resembling that found in experiments
\cite{tuteja2007} and simulations,\cite{savoy2012b} but with a
hydrophilic top layer (contact angle $\theta_\mathrm{top}=55^\circ$) combined with
a hydrophobic interior  ($\theta_\mathrm{in}=110^\circ$, see
Figure~\ref{fig:system}b).
To disentangle the effect of the geometry from that of the chemistry we
also simulate a purely hydrophobic surface
($\theta_\mathrm{in}=\theta_\mathrm{top}=110^\circ$) and a
purely hydrophilic one
($\theta_\mathrm{in}=\theta_\mathrm{top}=55^\circ$) with the same T shape.

RMD simulations are run at constant pressure and temperature for 
Lennard-Jones fluid and solids (see \emph{Supporting Information}). The
free energy profiles thus obtained are reported in
Figure~\ref{fig:system}c as a function of the filling fraction $\Phi$ of
the cavity for pressure close to two-phase coexistence, $\Delta P\approx
0$. 
The filling fraction is defined as $\Phi\equiv(z_W-z)/(z_W-z_C)$, where $z_W$ and $z_C$ are the number of
atoms inside the yellow box of Figure~\ref{fig:system}b corresponding to
the Wenzel and to the Cassie state, respectively.
The pressure difference $\Delta P\equiv P_l -P_v
-P_g$ is approximated as $\Delta P\approx P-P_v$, with $P_l$, $P_v$,
and $P_g$ the pressures in the liquid, vapor, and gas phases and $P$
the pressure of the barostat.\footnote{With the present definitions of the parameters, the atomistic and macro-
scopic results are in fair agreement and show the same trends; in order
to obtain a stringent quantitative comparison, however, one might need to
consider effects such as the liquid compressibility or the details of the baro-
stat. These more technical issues, which do not affect the present results,
are deferred to future work.} No other gas is
present in the simulations ($P_g=0$).

At $\Delta P\approx0$, the free energy profiles for the three chemistries 
exhibit two minima corresponding to the Wenzel and Cassie states. 
At negative $\Delta P$ a third metastable state emerges at large
$\Phi$, corresponding to the evaporated state.
Figure~\ref{fig:system}c shows that the Salvinia-like free energy profiles
(red) are, to a good approximation, a superimposition of the hydrophobic (in green, for $0<\Phi<1$) and the hydrophilic
ones (in blue, for $\Phi>1$). This explains the essential function of the
heterogeneous structure of the Salvinia: the hydrophobic interior \emph{stabilizes
the Cassie state} with respect to liquid intrusion (Cassie-Wenzel
transition), while the hydrophilic top \emph{hinders gas nucleation} (details in the
following). This is our main result, which at the same time clarifies in
quantitative terms the function of a complex biological structure, first
described by Barthlott and coworkers,\cite{solga2007,barthlott2010} and suggests
how to exploit it in the design of simpler bioinspired surfaces.

\begin{figure*}
	\centering
	\includegraphics[width=\textwidth]{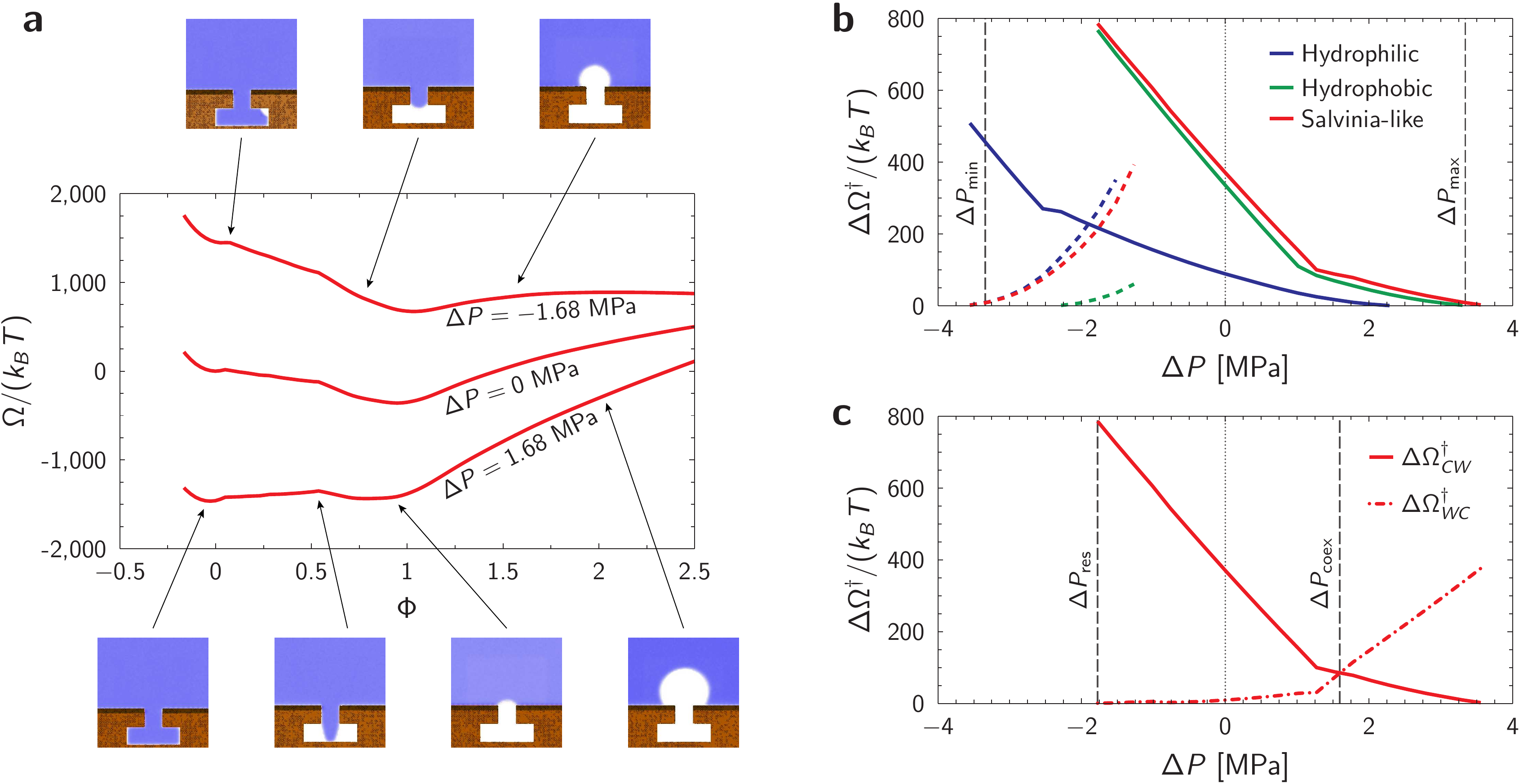}
	\caption{ 
		a) Free energy profiles for the Salvinia-like structure at different
		pressures; an arbitrary vertical shift is added for clarity. 
		The insets show the most probable configurations along the
		transition.
		b) Free energy barriers as a function of pressure for the three
		systems as computed from RMD simulations. Solid lines are used for $\Delta \Omega^\dag_{CW}$ and
		dashed ones for $\Delta \Omega^\dag_{Cv}$. The vertical lines are
		the macroscopic estimates for the spinodal pressures 
		$\Delta P_\mathrm{max}=2\gamma_{lv}/w$ and  $\Delta
		P_\mathrm{min}=-2\gamma_{lv}/w$.
		c)  Cassie-Wenzel (solid) and Wenzel-Cassie (dashed-dotted) free energy barriers as a
		function of pressure for the Salvinia-like structure. $\Delta P_\mathrm{res}$ is the spinodal
		pressure at which the Cassie state is spontaneously restored from the Wenzel
		one; $\Delta P_\mathrm{coex}$ is the coexistence pressure where the
		Cassie and Wenzel states are equiprobable (same free energy).
	\label{fig:barriers}}
\end{figure*}

Figure~\ref{fig:barriers}a addresses the effect of the pressure on the free energy profiles, which amounts to 
adding to $\Omega(\Phi)$ a term $\sim \Phi \Delta P$;\cite{Giacomello2012,giacomello2013}
this linear shift changes the location of the minima and determines the
stability of the Cassie state: for instance, increasing the pressure always favors the Wenzel state.  
At sufficiently large pressures, the Cassie minimum disappears and
$\Delta \Omega^\dag_{CW}\to0$: this is the \emph{spinodal} pressure
$\Delta P_\mathrm{max}$ for the Cassie-Wenzel transition, i.e., the
maximum pressure before the mechanical destabilization of
superhydrophobicity (mechanism $1$).
At $\Delta P<0$ (``negative pressures'') vapor bubbles tend to nucleate from the
T structure (``cavitation''). The thermodynamically stable vapor state is separated from the Cassie state
by the free energy barrier $\Delta \Omega^\dag_{Cv}$. 

Figure~\ref{fig:barriers}b reports $\Delta \Omega^\dag_{CW}$ and $\Delta
\Omega^\dag_{Cv}$ as a function of pressure for the three chemistries
considered. The barriers are typically hundreds of $k_BT$, accounting
for experimentally relevant metastabilities. Cavitation
 is favored by extreme negative pressures, which cause a reduction of $\Delta
\Omega^\dag_{Cv}$; this barrier vanishes at the Cassie-vapor spinodal
pressure $\Delta P_\mathrm{min}$  where the Cassie minimum disappears
(mechanism $4$).  The chemistry of the \emph{top layer} determines
$\Delta \Omega^\dag_{Cv}$, with the hydrophilic one having a much larger
barrier for cavitation.  On the other hand, $\Delta
\Omega^\dag_{CW}$ is large at negative pressures and monotonically
decreases with $\Delta P$; its value depends on  the chemistry of
the interior of the cavity, with the hydrophobic one having the
largest intrusion barrier.

Figure~\ref{fig:barriers}c reports the backward Wenzel-Cassie barrier $\Delta
\Omega^\dag_{WC}$. It is seen that the Cassie state can be \emph{res}tored at
pressures below the Wenzel-Cassie spinodal, $\Delta P_\mathrm{res}$,
where $\Delta \Omega^\dag_{WC} \to 0$. This result, which is not
captured by the macroscopic capillarity theory,\cite{giacomello2013} shows that the Wenzel state
can be ``reversible'', suggesting that superhydrophobicity can be
restored, albeit at negative $\Delta P$.

%%%%%%%%%%%%%%%%%%%%%%%%%%%%%%%%%%%%%%%%%%%%%%%%%%%%%%%%%%%%%%%

In the following, we go beyond the atomistic scale and
derive design principles of
general validity for superhydrophobic submerged surfaces.
Using the concepts
of classical capillarity, we first focus on the conditions of existence
of the superhydrophobic Cassie state and how these are affected by the
chemistry and topography of the surface texturing.  Then, we show how
the Salvinia-like structure is able to extend the range of pressures
where superhydrophobicity is stable.
The atomistic ``experiment'' and continuum
models are in qualitative agreement, confirming the general validity of
our design principles.

\begin{figure*}
	\centering
	\includegraphics[width=0.99\textwidth]{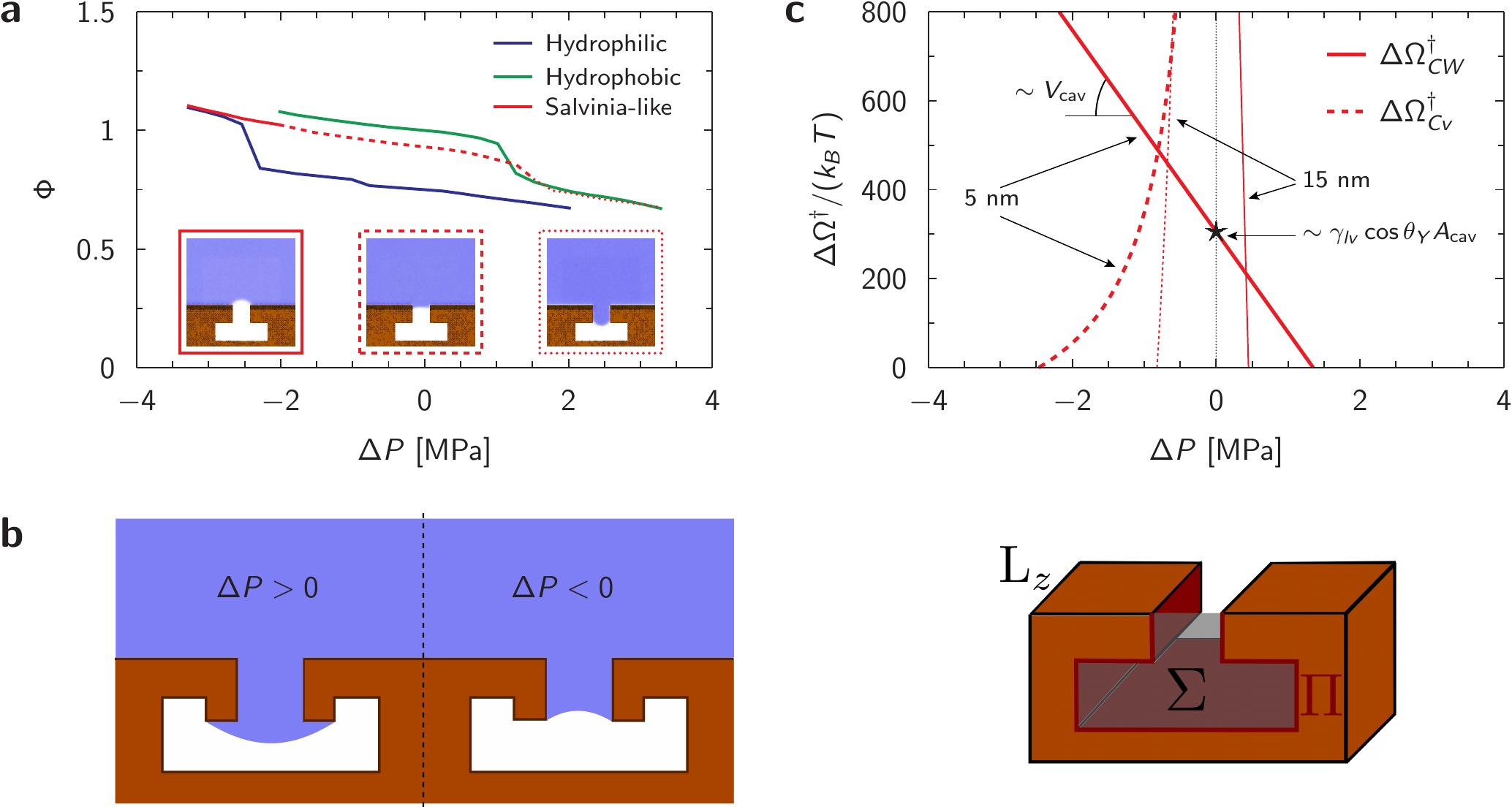}
	\caption{a) Filling level $\Phi$ at the Cassie minima of the free
		energy as a function of the pressure as computed from RMD. The
		Salvinia-like structure presents three different pinning regimes
		(insets): at the top corners ($\Delta P<0$), at the chemical contrast ($\Delta
		P\sim0$), and at the lower corners ($\Delta P>0$).
		b) Sketch of the serif T geometry, which can prevent intrusion even for
		liquids with low contact angles ($\theta_\mathrm{in}=\theta_\mathrm{top}\approx0$).
		c) Intrusion and nucleation free energy barriers as computed via
		approximate macroscopic expressions  (see \emph{Supporting Information}). 
		$\Delta \Omega^\dag_{CW}$ (solid lines) and $\Delta
		\Omega^\dag_{Cv}$ (dashed lines) are plotted for two systems with
		reference dimensions $w=5$~nm (thick lines) and $w=15$~nm (thin lines). 
		The volume of the cavity is given by $V_\mathrm{cav}=\Sigma
		\, L_z$ and its internal area by $A_\mathrm{cav}=\Pi \,L_z$.
		For the intrusion barrier we assume  $\Delta \Omega_{CW}^\dag \approx  -3\Delta P V_\mathrm{cav}/4  - 7
		\gamma_{lv}\cos\theta_\mathrm{in} A_\mathrm{cav}/9$; for the
		nucleation barrier  $\Delta \Omega^\dag_{Cv}\approx
		\pi\gamma_{lv}^2 L_z /\lvert \Delta P\rvert  -2\gamma_{lv} w L_z$. 
	\label{fig:minima}}
\end{figure*}

For the T geometry, the suspended
state is attained at the corners of the solid surface or at the chemical
contrast, which allow the \emph{pinning} of the contact line, Figure~\ref{fig:minima}a.  
In macroscopic terms this corresponds to the so-called
Gibbs' criterion,\cite{oliver1977} which prescribes that the range of
possible contact angles $\beta$ (Fig.~\ref{fig:system}b)
at a sharp corner or at a chemical contrast must be
included between the Young's angles approaching the discontinuity from
the two sides. On the T structure this pinning interval is
\begin{subequations}
		\label{eq:gibbs}
\begin{align}
\phi_1+\theta_\mathrm{top} -180^\circ
< & \beta< \theta_\mathrm{in} \label{betaCond:ext} \\
\theta_\mathrm{in}
< & \beta< 180^\circ -\phi_2+\theta_\mathrm{in} \label{betaCond:in}
\end{align}
\end{subequations}
with Equation~\ref{betaCond:ext} referring to the top
corners/chemical contrast and  Equation~\ref{betaCond:in} to the re-entrant ones
(for the definitions, see Figure~\ref{fig:system}b; for an extended
discussion, see the \emph{Supporting Information}).
	
From a mechanistic point of view, in the generic case of a periodic pattern of macroscopic structures, the
force balance at the Cassie state is given by (see, e.g., Ref.~\cite{checco2014}):
\begin{equation}
	\Delta P= -2 \gamma_{lv} \cos\beta\;	\frac{L}{A_\mathrm{mouth}}
	\label{eq:forcebalance}
\end{equation}
where $\gamma_{lv}$ is the liquid-vapor surface tension, $L$ is the
length of the contact line, and $A_\mathrm{mouth}$ is the liquid-vapor
area projected on the horizontal plane. For the T structure 
$L/A_\mathrm{mouth}=1/w$ with $w$ the width of the cavity
mouth; furthermore, the angle $\beta$ is limited by Equation~\ref{eq:gibbs}, which,
together with Equation~\ref{eq:forcebalance}, dictates the range of pressures where the Cassie
state exists: the minimum possible pressure is $\Delta P_\mathrm{min} \equiv \min_{\cos \beta} \left ( -2 \gamma_{lv}
\cos\beta\;	{L}/{A_\mathrm{mouth}} \right )$, while the maximum is
$\Delta P_\mathrm{max} \equiv \max_{\cos
\beta} \left ( -2 \gamma_{lv} \cos\beta\;	{L}/{A_\mathrm{mouth}} \right
)$. $\Delta P_\mathrm{min}$ and $\Delta P_\mathrm{max}$ are the spinodal pressures
for the Cassie state:  for $\Delta P \le \Delta P_\mathrm{min}$ the system
cavitates while for $\Delta P \ge \Delta P_\mathrm{max}$ the liquid
intrudes the cavities toward the Wenzel state.

For the Salvinia-like structure the intrusion and nucleation spinodals
are attained at $\beta_\mathrm{max}=180^\circ$ and
$\beta_\mathrm{min}=0^\circ$ which plugged into Equation~\ref{eq:forcebalance}
yield $\Delta P_\mathrm{max}=2\gamma_{lv}/w$ and $\Delta
P_\mathrm{min}=-2\gamma_{lv}/w$, respectively.  A crucial feature of the
Salvinia-like structure, therefore, is that both spinodals are not
explicitly dependent on the chemistry of the surface; the chemistry,
together with the re-entrant topography of the surface, only ensures that 
$\beta_\mathrm{max}=180^\circ$ and $\beta_\mathrm{min}=0^\circ$ are
within the pinning interval of Equation~\ref{eq:gibbs}.  Importantly,
these values also maximize the range of pressures where the
superhydrophobic state exists for a given $w$. 
In the chemically homogeneous cases, instead, this pressure range
is smaller and explicitly depends on the chemistry (see \emph{Supporting Information}).

Summarizing, in order to realize these
optimal conditions for submerged superhydrophobicity, the geometry 
of the cavity mouth should be combined
with the chemistry in such a way that 
$180^\circ - \phi_2 + \theta_\mathrm{in} \geq 180^\circ$
and $\phi_1 + \theta_\mathrm{top} -  180^\circ \leq 0^\circ$.
More complex geometries, such as the
doubly re-entrant ``serif T'' ($\phi_1=90^\circ$ and $\phi_2\approx 0^\circ$),\cite{hensel2013} can be designed in
order to repel liquids with low contact angles
($\theta_\mathrm{in}=\theta_\mathrm{top}\approx 0^\circ$).\cite{liu2014}
In this case, the meniscus is pinned at the innermost
corner for $\Delta P>0$ (see Figure~\ref{fig:minima}b and \emph{Supporting Information}).

%%%%%%%%%%%%%%%%%%%%%%%%%%%%%%%%%%%%%%%%%%%%%%%%
Based on Equation~\ref{eq:landau} and \ref{eq:rate} we now discuss the
stability of submerged superhydrophobicity.  As compared to
drop-repellent surfaces, typical submerged applications require
the superhydrophobic state to survive for longer times and at
comparatively larger pressures.  The thermodynamic stability of the
Cassie state -- i.e., Cassie being the absolute minimum of
$\Omega(\Phi;P,T)$ -- is often invoked in order to obtain such
``robust'' superhydrophobicity.  
This criterion usually requires overly tall and fragile structures and
is redundant since the duration $\tau(P,T)$ of a metastable Cassie state is
typically much larger than the experimental timescale.
For our Salvinia-like nanostructure, Equation~\ref{eq:rate} --
assuming conservatively molecular timescales for the prefactor,\cite{eyring1935} 
$\tau_0 = h/(k_B T) \approx 10^{-13}$~s, where $h$ is
Planck's constant, and $\Delta \Omega^\dag\sim 100\; k_B T$
(Fig.~\ref{fig:barriers}b) -- predicts that the lifetime of the Cassie state  exceeds the age of the universe. 
In other words, if the free energy barriers are sufficiently large the
superhydrophobic  state -- stable or metastable -- is robust and
mechanisms $2$ and $3$ of gas loss are in practice inhibited.

The typical trend of the free energy barriers with the characteristic size $w$
of the surface texturing is shown in Figure~\ref{fig:minima}c: increasing the size of the cavity decreases
$\Delta P_\mathrm{max}$ and dramatically increases the dependence of
$\Delta \Omega_{CW}^\dag$ on $\Delta P$; the effect on cavitation is
similar.
Thus, the thermally activated breakdown of superhydrophobicity
(mechanisms $2$ and $3$ of gas loss) becomes important only in the vicinity of the
spinodals, where the barrier is of the order of the
thermal energy $k_BT$. The amplitude of this region rapidly shrinks
with the size of the structures (see Figure~\ref{fig:minima}c).

%%%%%%%%%%%%%%%%%%%%%%%%%%%%%%%%%%%%%%%%%%%%%%%%

In summary, atomistic rare event simulations have unraveled the Salvinia
paradox: a re-entrant geometry,
together with a hydrophobic interior, improves the stability of gas
pockets against liquid intrusion and contaminants, while the hydrophilic
top surface hinders the nucleation and coalescence of bubbles.  This
natural paradigm reveals two simple design principles for engineering
submerged surfaces: the pinning interval can be tuned via the chemistry
and surface topography (Equation~\ref{eq:gibbs}) and the range of
positive and negative pressures where superhydrophobicity is
(meta)stable can be controlled via the size of the cavity mouth
(spinodal pressures). Smaller structures typically correspond to larger
superhydrophobic pressure ranges; however, the range of pressures over
which the thermally activated breakdown of superhydrophobicity is
possible broadens for smaller structures, eventually limiting this
``shrinking'' strategy.

%%%%%%%%%%%%%%%%%%%%%%%%%%%%%%%%%%%%%%%%%%%%%%%%

\subsection*{Supporting Information}
Supporting Information is available from the Wiley Online Library or
from the author.

\subsection*{Acknowledgements}
The research leading to these results has received funding from
the European Research Council under the European Union's
Seventh Framework Programme (FP7/2007-2013)/ERC Grant agreement n. [339446].  
We acknowledge PRACE for awarding us access to resource
FERMI based in Italy at Casalecchio di Reno.

\bibliography{biblio}

%Fig3
%

\end{document}